\documentclass[12pt]{article}
\usepackage{pic04}
\usepackage{hyperref}
\usepackage{url}
\usepackage{graphicx}

\newcommand{\met}{\mbox{${\hbox{$E$\kern-0.6em\lower-.1ex\hbox{/}}}_T$}}
\newcommand{\rmscr}[1]{\ensuremath{\mbox{\scriptsize{#1}}}}
\newcommand{\qqbar}{\ensuremath{\mathrm{q\bar{q}}}}
\newcommand{\ttbar}{\ensuremath{\mathrm{t\bar{t}}}}
\newcommand{\ppbar}{\ensuremath{p\bar{p}}}
\newcommand{\invpb}{\ensuremath{\mbox{pb}^{-1}}}
\newcommand{\invfb}{\ensuremath{\mbox{fb}^{-1}}}
\newcommand{\mt}{\ensuremath{m_{\mathrm{t}}}}
\newcommand{\GeVcsq}{\ensuremath{\mbox{GeV}/c^{2}}}

\begin{document}

\title{\bf TEVATRON RESULTS ON TOP QUARK PHYSICS.}
\author{
Andy Hocker \\
for the CDF and D0 collaborations        \\
{\em University of Rochester, Rochester NY, 14627-0171}}
\maketitle

%
%
\begin{figure}[h]
\begin{center}
%
%
%
%
\vspace{4.5cm}
\end{center}
\end{figure}

\baselineskip=14.5pt
\begin{abstract}
The CDF and D0 collaborations at Fermilab's Tevatron \ppbar\ collider have in place
an extensive program to measure fundamental properties of the top quark.  Recent
results from Run~I ($\sqrt{s}~=~1.8$~TeV) and Run~II ($\sqrt{s}~=~1.96$~TeV)
on the top quark's production, mass, and decays are presented here.  All results
are consistent within their uncertainties with the Standard Model expectations for
the top quark. 
\end{abstract}
\newpage

\baselineskip=17pt

\section{Introduction}
After the discovery of the top quark in Run~I of Fermilab's Tevatron collider~\cite{disco},
the CDF and D0 collaborations switched from ``discovery mode'' to ``measurement mode,''
seeking to fully characterize the properties of this newest fundamental particle.
However, the limited size of the Run~I data sample ($\sim 100~\invpb$) left most
measurements with large statistical uncertainties.

Run~II of the Tevatron is expected to yield 4--8 \invfb\ and therefore
a far more detailed and necessary study of the physics of the top sector.  By far the most
intriguing aspect
of the top quark is its mass; at 178 \GeVcsq\ (about the same as an entire gold atom)
it weighs in right at the electroweak symmetry breaking (EWSB) scale, which could be
a clue to the origin of EWSB.  Its large mass also leads to a lifetime shorter than
the hadronization time scale, meaning it decays as a free quark and therefore provides
the opportunity to probe heretofore-inaccessible bare quark properties such as spin and
charge.  Its production and decay can provide direct contact with the CKM matrix element
$V_{tb}$ and the study of the electroweak interaction at high energy.  Finally, its
energetic multi-body final states are often an important background in searches for
new heavy particles (the Higgs boson being the canonical example).

\section{Top Production and Decay at the Tevatron}
\label{sec:proddecay}
\begin{figure}[htb]
\centerline{
  \includegraphics[width=0.6\textwidth]{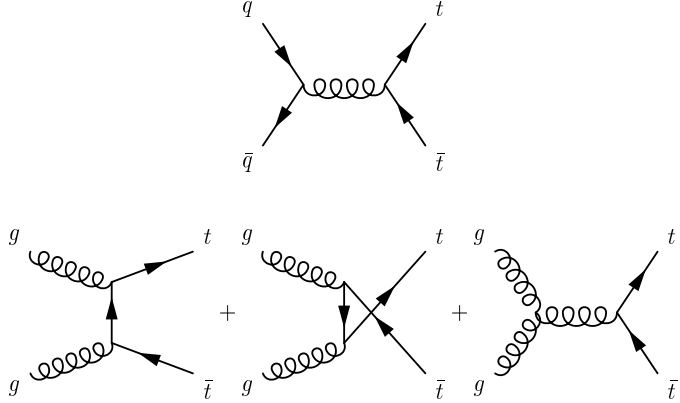}
}
 \caption{\it
      Leading Feynman diagrams for top pair production at the Tevatron.
    \label{fig:ttprod} }
\end{figure}

The main mechanism for producing top quarks at the Tevatron is pair production via
the strong interaction (electroweak production will be discussed in Section~\ref{sec:singlet}).
\qqbar\ annihilation
accounts for about 85\% of \ttbar\ production with the remainder taken up by
gg processes (see Figure~\ref{fig:ttprod}).  Calculations of the
$\ppbar \rightarrow \ttbar$ cross section~\cite{tt_xsec} for $\sqrt{s}~=~1.96$~TeV and
\mt~=~175~\GeVcsq\ result in $\sigma~=~6.7^{+0.7}_{-0.9}$~pb.  This represents an increase
of approximately 30\% with respect to Run~I due to the larger center-of-mass energy.

Within the Standard Model the top quark decays to a bottom quark and a W boson nearly 100\%
of the time.  Therefore, the final states that characterize \ttbar\ production are determined
by whether the W bosons decay hadronically or leptonically.\footnote{W decays to taus
are only considered leptonic if the tau itself decays leptonically.  Hadronic tau decays
present their own difficulties and are usually considered separately.}  The main final states are
\begin{itemize}
  \item {\bf Dilepton} ($\ell\nu\mathrm{b}\ell\nu\mathrm{b}$):  The branching fraction for this final
state is only about 7\%, but the presence of two energetic leptons and large missing transverse energy
(\met) from the undetected neutrinos yields a clean event sample even without requiring explicit
b-jet identification (``b-tagging'').  The main backgrounds for this channel are Drell-Yan production,
diboson production, and fake leptons.
  \item {\bf Lepton + jets} ($\ell\nu\mathrm{b}jj\mathrm{b}$):  With a branching fraction of about 35\%,
this is something of a ``golden mode'' for \ttbar\ production.  The main background is the production
of W bosons with associated jets, but after b-tagging good purity can be obtained.
  \item {\bf All-hadronic} ($jj\mathrm{b}jj\mathrm{b}$):  This channel has the largest branching
fraction (44\%), but suffers from an enormous QCD multijet background.  b-tagging is essential
in this channel; however the purity of this final state is far worse than the channels with a lepton.
\end{itemize}

\section{Detectors and Data Samples}
The final states outlined above determine the requirements for a detector expected to do top physics.
Good electromagnetic calorimetry and muon detection are required with as much reach in pseudorapidity
($\eta$) as possible.  Well-calibrated hadron calorimetry covering as much solid angle as possible is
necessary for good \met\ and jet energy resolution, and efficient b-tagging requires precision tracking
of charged particles.  All these considerations entered into the upgrade of the CDF and D0 detectors
for Run~II.  In addition to complete upgrades of their trigger and data acquisition systems to cope with
higher collision rates, both detectors augmented their central tracking systems with new silicon microstrip
detectors and improved their forward muon identification capabilities.  Also key to the top physics program
were CDF's new forward calorimeter system for improved electron identification, and a 2~T spectrometer magnet
for charged particle momentum measurement at D0.

The Tevatron became operational in March 2001, and at conference time both experiments had about
400 \invpb\ of collision data on tape.  Many of these data were obtained in 2004; the results presented
here stem from data taken prior to a long accelerator shutdown at the end of 2003.  The amount of data
used typically ranges from about 150 to 200 \invpb, depending on subdetector quality requirements that
vary from analysis to analysis.

\section{Cross Section Measurements}
Measuring the production cross section is a first step in any top physics program, as it establishes
a baseline event selection and determines the amount of signal and background present in the sample.
It is also an interesting measurement in its own right, as it provides a test of the Standard Model
predictions for top production via QCD.  Deviations from the SM could signal an exotic production
mechanism (such as a heavy \ttbar\ resonance) or a ``contamination'' of the \ttbar\ sample from new
physics processes.  CDF and D0 have measured the \ttbar\ cross section in a number of channels in a
number of different ways; a brief overview of the techniques is given below.

\begin{figure}[htb]
\centerline{
  \includegraphics[width=0.6\textwidth]{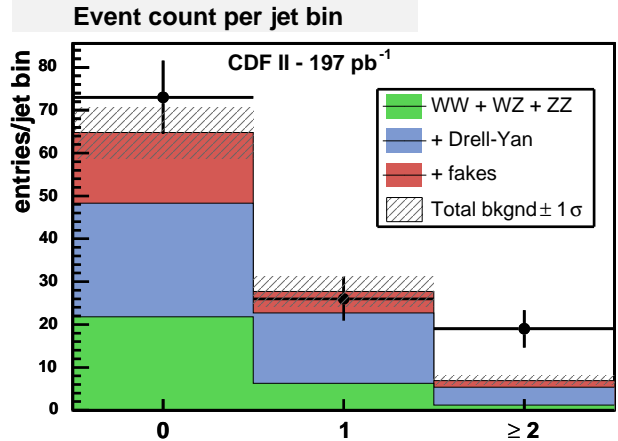}
}
 \caption{\it
      Expected and observed number of events per jet multiplicity bin for CDF events with
an identified lepton, an isolated track, and large missing transverse energy.
    \label{fig:ltrk} }
\end{figure}

The cross section measurement in the dilepton channel is traditionally done as a counting experiment
using events with two energetic leptons, large \met, and at least two jets.  In order to increase
acceptance, one can sacrifice some of the high purity of the dilepton channel by relaxing identification
requirements on one of the leptons; one CDF analysis allows any isolated track to be considered as
a ``second'' lepton.  Figure~\ref{fig:ltrk} is a typical illustration of the counting method; the
\ttbar\ signal populates the high jet multiplicity bin and the lower multiplicities are used to cross-check
background predictions.  In addition to the counting experiments, CDF has employed a new technique
wherein all events with two leptons are fit to the expected distributions from signal and background
in the (\met,$N_{\rmscr{jets}}$) plane, resulting in greater statistical power.

A variety of methods for measuring the cross section are available in the lepton+jets channel.
By exploiting the long lifetime of bottom hadrons one can identify a b-quark jet by
reconstructing a displaced secondary vertex or by counting tracks in a jet with large impact
parameter.  One can also look for ``soft'' leptons resulting from semileptonic b decays.  The
efficiency for tagging a jet in a \ttbar\ event is typically about 50\%; by requiring at least
one tag a counting experiment can be done to extract the cross section.  In lieu of requiring
a tag, one can also exploit the large mass of the top quark; the jets in \ttbar\ decay tend
to be more energetic than those in the dominant W+jets background.  By fitting the data to
a discriminant variable such as the scalar $E_{T}$ sum of all objects in the event or to a
multivariate kinematic discriminant such as the output of an artificial neural network, one
can determine the fraction of \ttbar\ in the W+jets sample and thus the cross section.

The background-swamped all-hadronic channel requires one to exploit both the top quark's
large mass and the presence of b quarks.  In this channel CDF measures the cross section
by counting the number of excess b-tags in a sample of events with at least six jets that
are selected to have topologies indicative of the production of heavy t quarks.  D0 extracts
the cross section by fitting b-tagged multijet data to the output of an artificial neural
network constructed from discriminant topological variables.

\begin{figure}[htbp]
  \centerline{\hbox{ \hspace{4.0cm}
    \includegraphics[width=0.4\textwidth]{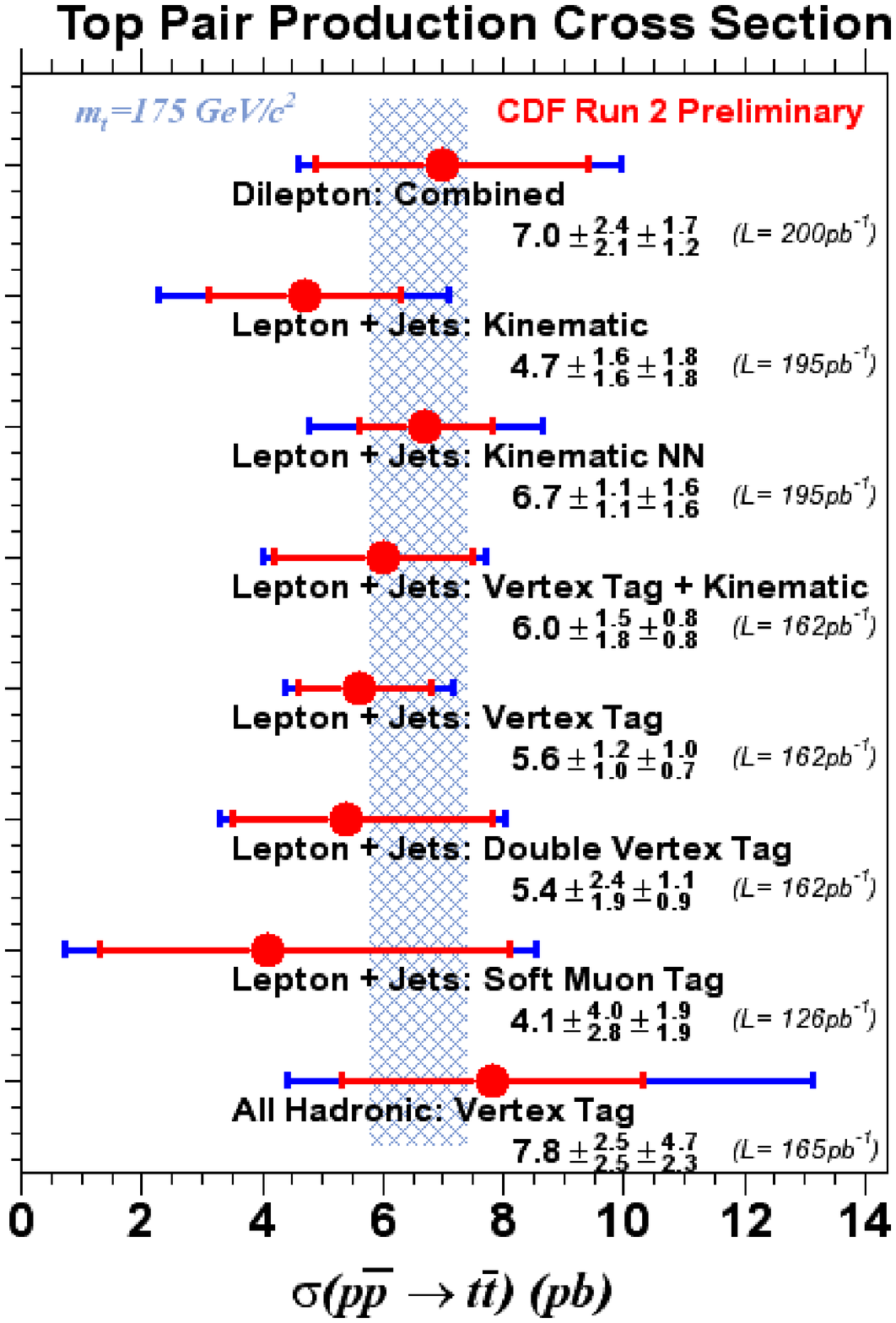}
    \hspace{0.3cm}
    \includegraphics[width=0.6\textwidth]{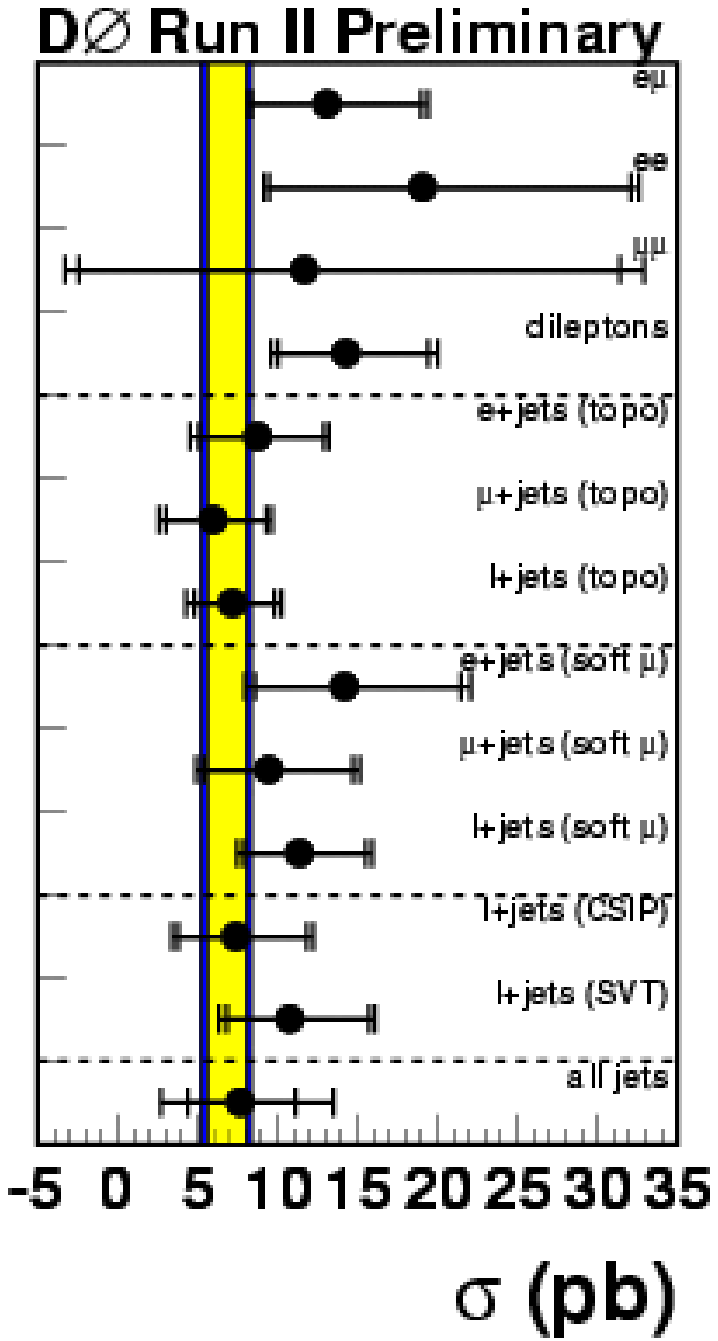}
    }
  }
 \caption{\it
      Summary of top pair production cross section measurements from CDF (left)
and D0 (right).
    \label{fig:xsec_results} }
\end{figure}

The results of the myriad cross section measurements are shown in Figure~\ref{fig:xsec_results}.
Note that they are consistent across the different decay channels, between the two
experiments, and with the SM expectation.

\section{Mass Measurements}
\begin{figure}[htb]
\centerline{
  \includegraphics[width=0.6\textwidth]{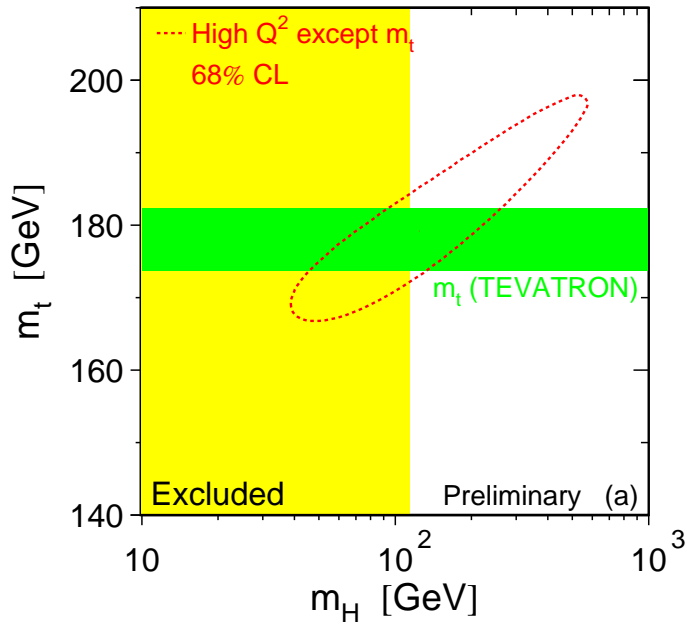}
}
 \caption{\it
      Results from global EW fits in the ($m_{H},m_{t}$) plane.  The yellow area
is excluded by direct Higgs searches at LEP2; the green area is the 1$\sigma\ m_{t}$
band determined from Run~I top mass measurements.  The elliptical region is allowed at
68\% CL according to the precision EW measurements (excluding $m_{t}$).
    \label{fig:mhmt} }
\end{figure}

A precise measurement of the top quark mass is critical for experimentally testing
the self-consistency of the Standard Model.  Along with other precision electroweak
measurements, it serves to constrain the possible values of the Higgs boson mass, as
illustrated in Figure~\ref{fig:mhmt}.  Current global EW fits place a 95\% CL upper bound
on the Higgs mass of 237~\GeVcsq~\cite{lepew}; the Run~II goal is to further constrain
the mass range by measuring the top mass to an uncertainty of 2--3~\GeVcsq.

The current EW fits use as an input a new Run~I average top mass value of $178.0\pm4.3$~\GeVcsq.
This new value is largely driven by a new analysis of D0's Run~I lepton+jets data
that yielded an unprecedented single-measurement precision for the top mass~\cite{d0mass}.
This analysis assigned a mass-dependent probability to each event based on the event's
kinematical compatibility with the leading-order mass-dependent matrix element for
\ttbar\ production and decay.  The probability-versus-mass curves for each event
were combined (the sharpness of the curve effectively weighting each event) and the
joint probability maximized to yield
$\mt~=~180.1\pm3.6\mbox{(stat.)}\pm3.9\mbox{(syst.)}$~\GeVcsq.  The increase
in statistical power using this technique is equivalent to a factor-of-2.4 larger dataset.

\begin{figure}[htb]
\centerline{
  \includegraphics[width=0.6\textwidth]{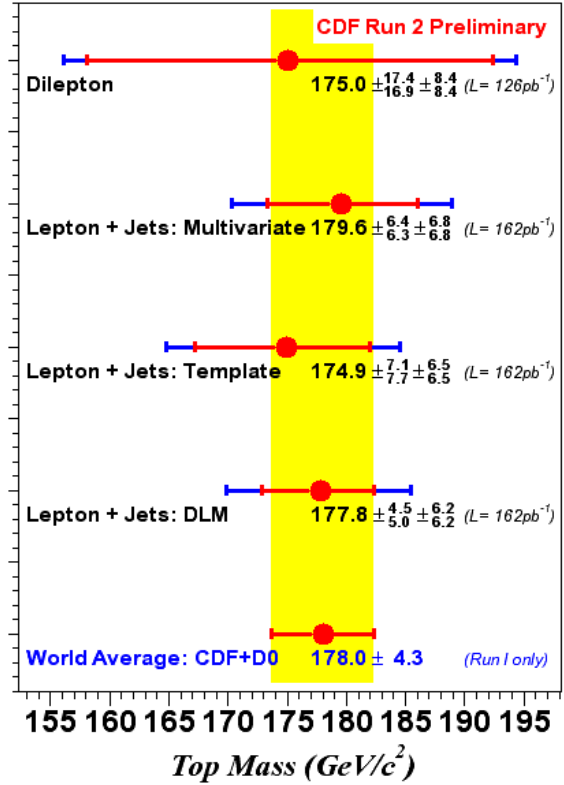}
}
 \caption{\it
    Summary of Run~II mass measurements.
    \label{fig:masssum} }
\end{figure}

Measurements from large Run~II datasets will soon be eclipsing the Run~I results, however.
CDF has performed several measurements of the top mass in the dilepton and lepton+jets
channels using Run~II data.  These measurements have been done using traditional ``template''
methods, where one mass is reconstructed per event and the resulting mass distribution
compared against template distributions from simulated \ttbar\ events of varying masses.
New techniques have also been employed, such as using multivariate templates and weighting
events according to the probability for the chosen jet-parton assignment to be correct.
An analysis based on probabilities formed from the \ttbar\ matrix element, similar to
the technique used for the D0 Run~I data, yields the best measurement of the top mass
in Run~II: $\mt~=~177.8^{+4.5}_{-5.0}\mbox{(stat.)}\pm6.2\mbox{(syst.)}$~\GeVcsq.
The Run~II mass measurements are summarized and compared with the Run~I average in
Figure~\ref{fig:masssum}.  Nearly all top mass measurements are quickly becoming
limited by the systematic uncertainty, the largest source of which is the uncertainty
on the jet energy scale.  Greatly improved understanding of the calorimeter response to
jets will be the key to achieving a high-precision measurement of the top mass in Run~II.

\section{Properties of Top Decays}
As mentioned in Section~\ref{sec:proddecay}, top is expected to decay to Wb nearly
100\% of the time.  This assumption can be checked by examining various decay
characteristics in detail; deviations from expectations could signal the presence
of new particles in top decays, or that what is in the data isn't really ``top'' at all!

\subsection{Top decays to taus}
The decay $\mathrm{t}\rightarrow\mathrm{Wb}\rightarrow\tau\nu\mathrm{b}$ is all third-generation
and is an excellent place for new physics to appear; for example supersymmetric models
with a light charged Higgs and large $\tan{\beta}$ could result in a large
$\mathrm{t}\rightarrow\mathrm{Hb}\rightarrow\tau\nu\mathrm{b}$ decay rate.  However,
most taus (65\%) decay hadronically and are quite difficult to distinguish from a
low-multiplicity parton jet.  CDF has looked for top decays to hadronically-decaying
taus in \ttbar\ events where one top decays to an electron or muon.  Two events are observed
in 193~\invpb\ with 2.4 events expected from SM top and background.  The branching
ratio for top into tau is therefore determined to be less than 5 times the SM expectation
at 95\% CL.

\subsection{Top decays to Xb}
Consider the possibility that, instead of a W, tops decay into some particle X with branching
ratio $\beta$, and that X decays hadronically.\footnote{An example of such a particle would
be a low-mass charged Higgs in certain SUSY scenarios.}  In that case the \ttbar\ cross sections
measured in the dilepton and lepton+jets channels would not agree.  Therefore we can test for this
possibility by taking the ratio $R$ of the dilepton and lepton+jets cross section;
a value significantly different from unity could signal the presence of X.  This has been
done, using CDF cross section results older than those presented in these proceedings, yielding
$R~=~1.45^{+0.83}_{-0.55}$.  Assuming the efficiency to detect X is the same as that for W,
one can then set a 95\% CL upper limit on $\beta$ at 0.46.  Similarly, one can set a limit on
$\beta^{\prime}$, the branching fraction of top into a leptonically-decaying X, at 0.47.

\subsection{Top decays to light quarks}
Assuming three-generation unitarity, the CKM matrix element $|V_{tb}|~\approx~0.999$, implying that
the ratio $b~=~BR(\mathrm{t}\rightarrow\mathrm{Wb})/BR(\mathrm{t}\rightarrow\mathrm{Wq})$
is nearly unity.  One can use the ratio of single-tagged top events to double-tagged top events
to measure $b$ if the b-tagging efficiency is known.  CDF performs this measurement by dividing
its lepton+jets sample into events with exactly three jets and four or more jets, and subdividing
into single- and double-tagged events.  These subsamples have different sensitivities to the
product $b\epsilon_{b}$, where $\epsilon_{b}$ is the efficiency to tag a b-quark jet.  The
number of observed events in each subsample is used to form a likelihood as a function of
$b\epsilon_{b}$.  The most likely value of $b\epsilon_{b}$ is then divided by the b-tagging
efficiency to extract $b~=~0.54^{+0.49}_{-0.39}$.  The large uncertainties on this result
can be readily improved by adding the information available in the zero-tagged data, as well
as adding the dilepton channel.

\subsection{Top dilepton kinematics}
\begin{figure}[htbp]
  \centerline{\hbox{ 
    \includegraphics[width=0.45\textwidth]{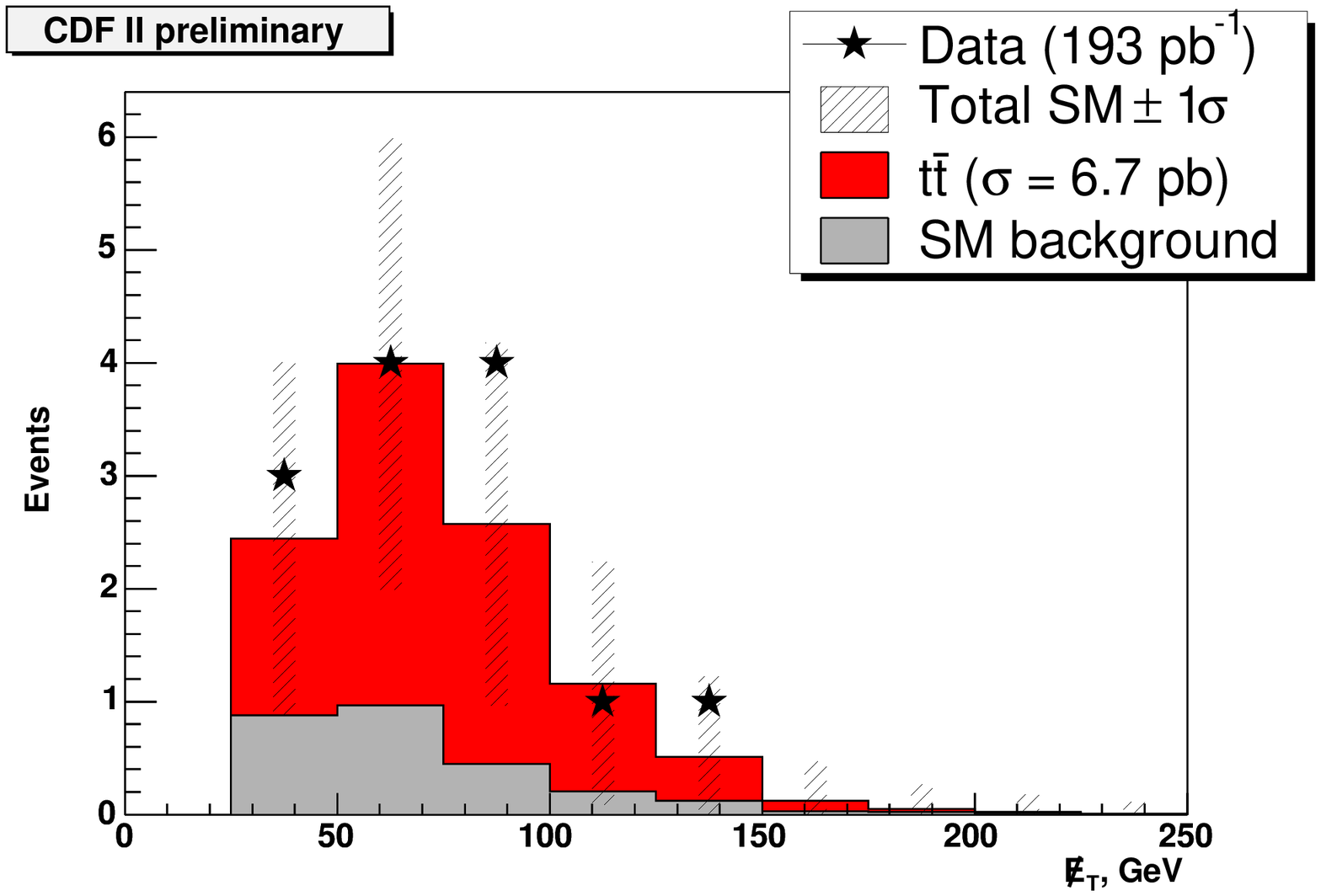}
    \hspace{0.3cm}
    \includegraphics[width=0.45\textwidth]{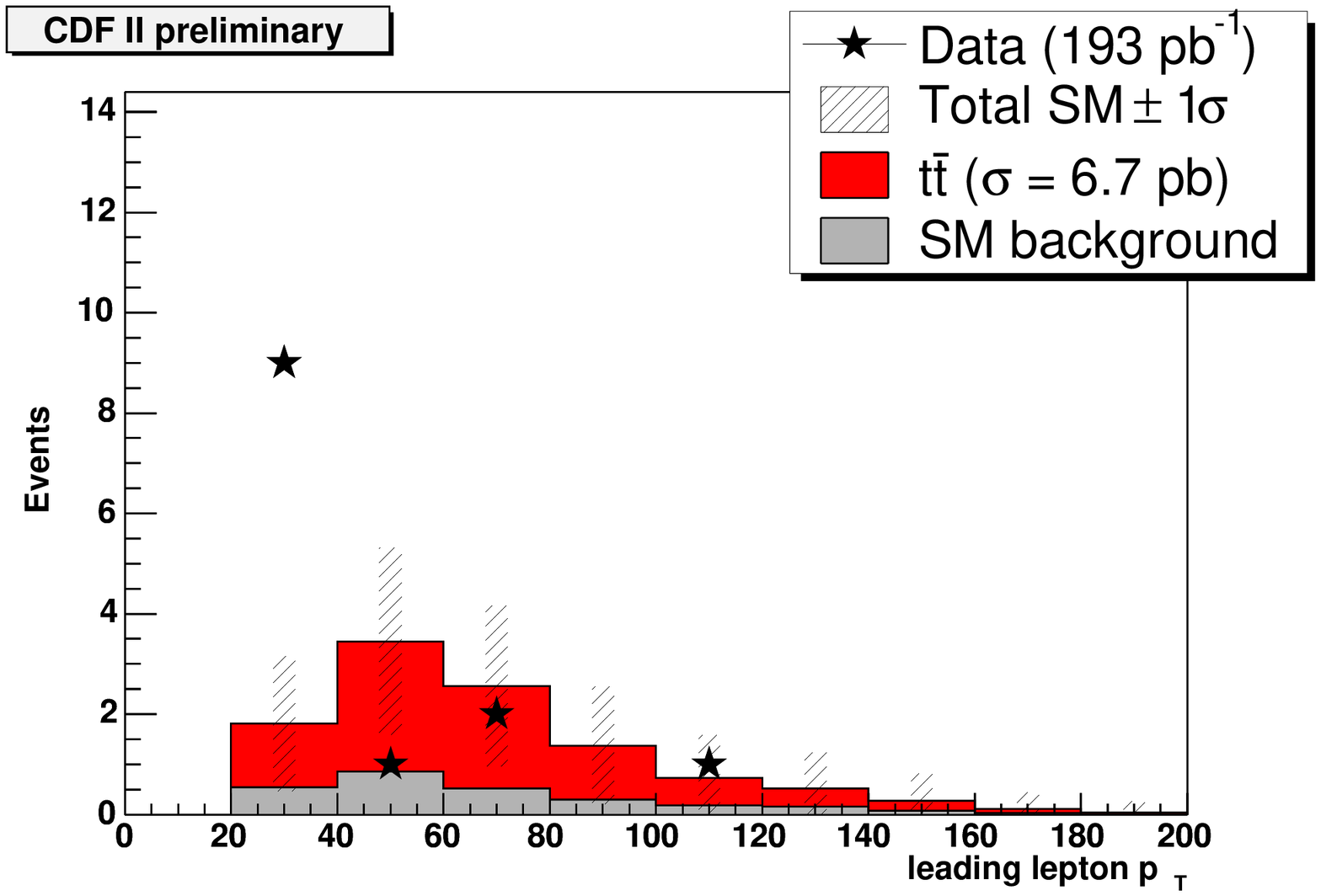}
    }
  }
 \caption{\it
    Expected and observed \met\ and leading lepton $p_{T}$ distributions in the Run~II
CDF dilepton sample.  The hatched regions indicate the 1$\sigma$ uncertainty on the
expectation in each bin.
    \label{fig:dilkin} }
\end{figure}

Several events in the Run~I top dilepton sample were characterized by values of \met\
and lepton $p_{T}$ considerably larger than the expectations from SM top.  In fact, it
was suggested that the kinematics of these events were better described by cascade decays
of heavy squarks~\cite{bandh}.  In order to prepare for a possible continuation of this
effect in Run~II, CDF developed an {\it a priori} method to assess the consistency of
the dilepton kinematics with the SM.  A four-variable phase space was chosen; two of
these variables are shown in Figure~\ref{fig:dilkin}.  No effects similar to those seen
in Run~I were observed; the data are quite consistent with expectations except for
(ironically) a mild excess at {\it low} lepton $p_{T}$.  The probability of the data's
consistency with the SM in this phase space is in the range 1.0--4.5\%; a detailed
look at the low-$p_{T}$ events that drive this result point to a fluctuation of SM
top as a likely interpretation.

\subsection{W helicity measurements}
\begin{figure}[htb]
\centerline{
  \includegraphics[width=0.6\textwidth]{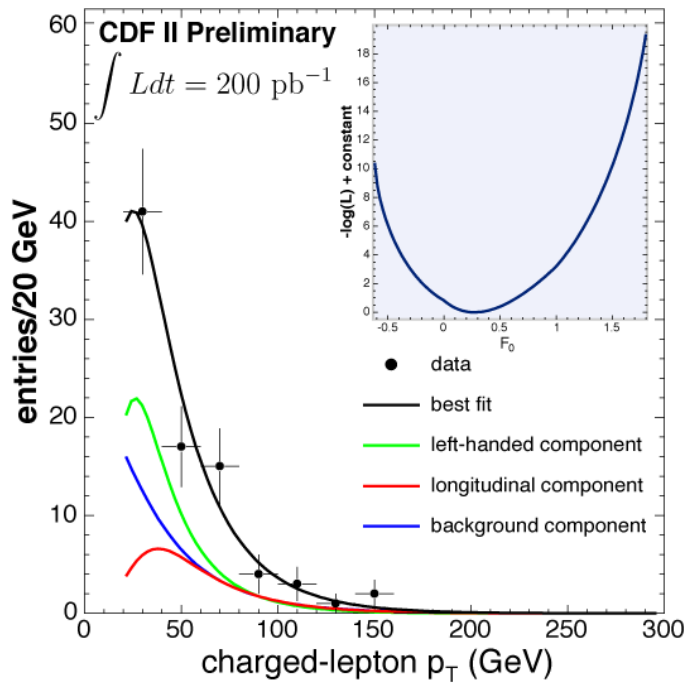}
}
 \caption{\it
    Lepton $p_{T}$ spectrum for CDF dilepton and lepton+jets events.  The inset shows the
fit likelihood as a function of $F_{0}$.  The blue line shows the (fixed) background
component; the red and green lines show the longitudinal and left-handed components returned
by the fit.
    \label{fig:whel} }
\end{figure}

Measuring the helicity of the W in top decays allows one to test the $V-A$ structure
of the electroweak interaction at high energies.  Due to angular momentum conservation,
tops decay only into left-handed (negative helicity) or longitudinally-polarized (zero helicity)
W bosons.  The fraction of longitudinally-polarized W bosons ($F_{0}$) is determined by
the ratio of the top and W masses; the SM prediction is $F_{0}~=~0.70$.  The helicity
of the W manifests itself in the kinematics of its decay products.  There are several ways
to exploit this; for example, the different helicity amplitudes give rise to different
distributions in $\cos{\theta^{*}}$, where $\cos{\theta^{*}}$ is the angle in the W rest
frame made by the charged lepton direction and the W flight direction.  D0 has measured
$F_{0}$ from Run~I lepton+jets data using a technique similar to that used to measure the
top mass; for each event an $F_{0}$-dependent probability is formed based on the event's
compatibility with the leading order matrix element expressed as a function of $F_{0}$.
The combined probability for the data sample is then maximized to extract
$F_{0}~=~0.56\pm0.31$.  CDF has measured $F_{0}$ in Run~II exploiting the tendency for
the charged leptons to be thrown parallel to right-handed W bosons and anti-parallel to
left-handed W bosons.  This method allows one to use both the lepton+jets and dilepton
samples in a fit to the lepton $p_{T}$ spectrum; see Figure~\ref{fig:whel}.  The measured $F_{0}$ is
$0.27^{+0.35}_{-0.24}$, about $1\sigma$ below the SM value.  One expects this result
given the observed excess of low-$p_{T}$ leptons in the dilepton channel noted above.

\section{Search for single top}
\label{sec:singlet}
\begin{figure}[htbp]
  \centerline{\hbox{ 
    \includegraphics[width=0.2\textwidth]{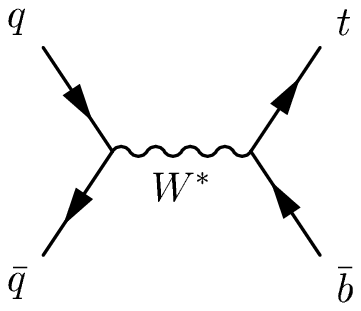}
    \hspace{1.0cm}
    \includegraphics[width=0.45\textwidth]{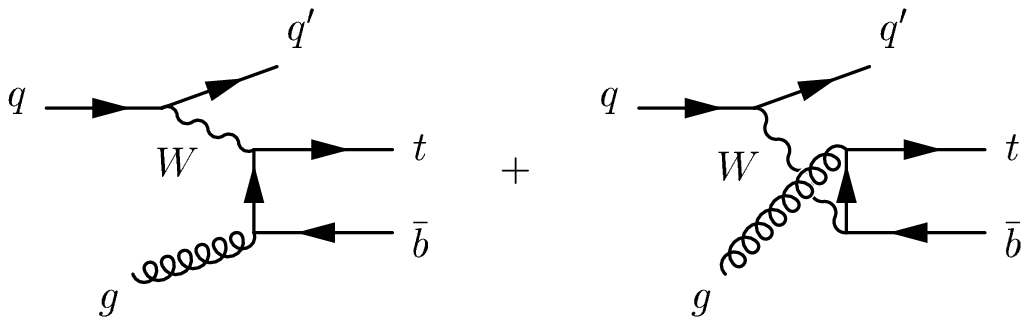}
    }
  }
 \caption{\it
    Feynman diagrams for the electroweak production of single top at the Tevatron.
    \label{fig:ewtop} }
\end{figure}

Top quarks can also be produced singly at the Tevatron via the electroweak
interaction through the $s$- and $t$-channel processes shown in Figure~\ref{fig:ewtop}.
This production mechanism is of considerable interest since the rate is directly
proportional to $|V_{tb}|^{2}$.
However, the lepton+jets final states that characterize single top have lower jet
multiplicities than for \ttbar, and hence considerably more W+jets background.  That
coupled with the smaller cross section for single top (a factor of 2--3 less than
\ttbar~\cite{singlet_xsec}) has precluded observation of single top production so far.
The cross section could be considerably enhanced, however, by the existence of a heavy
$\mathrm{W}^{\prime}$ or an anomalous tWb coupling.  Therefore searches for single
top are well underway at Run~II.

\begin{figure}[htbp]
  \centerline{\hbox{ 
    \includegraphics[width=0.45\textwidth]{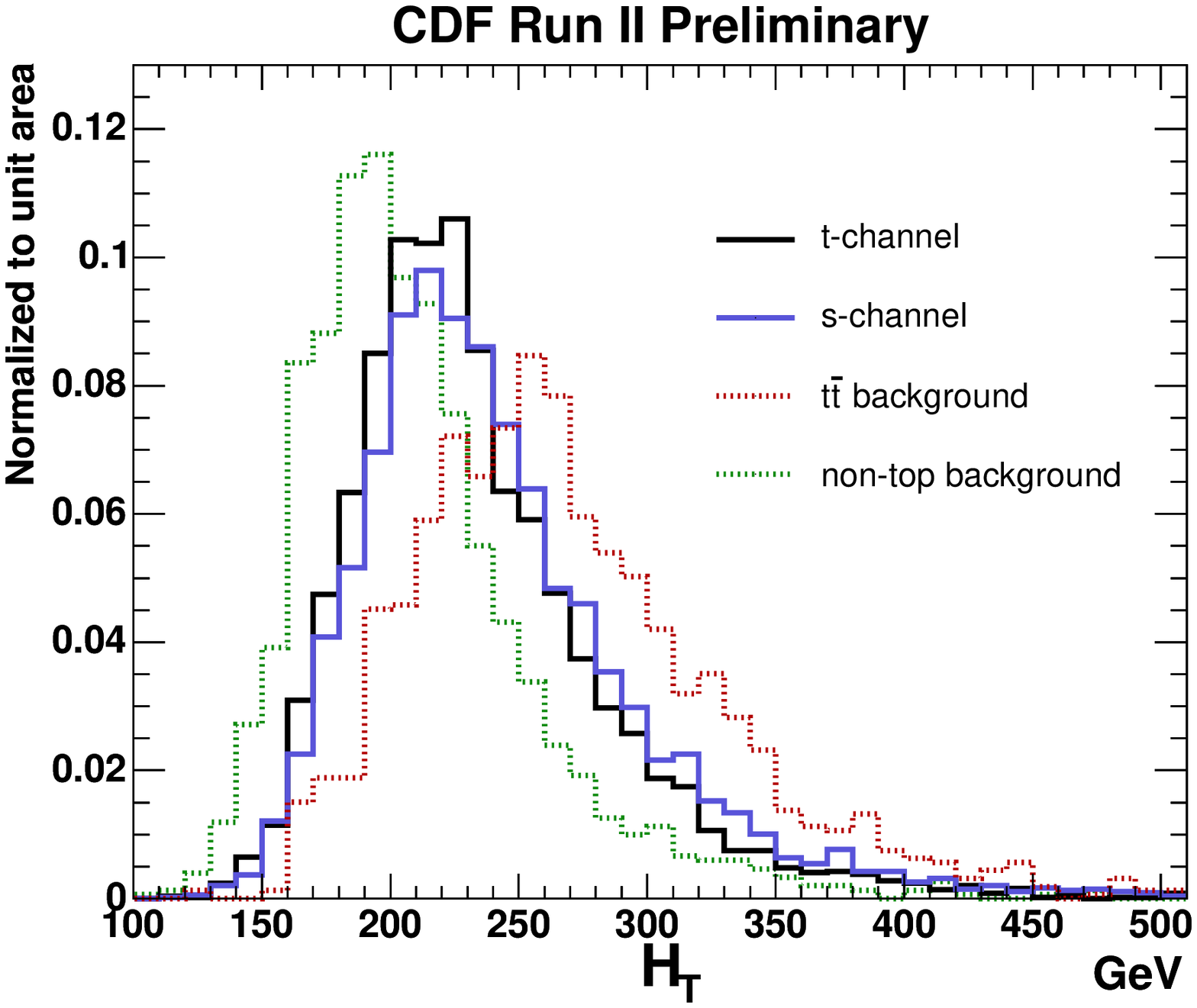}
    \hspace{0.3cm}
    \includegraphics[width=0.45\textwidth]{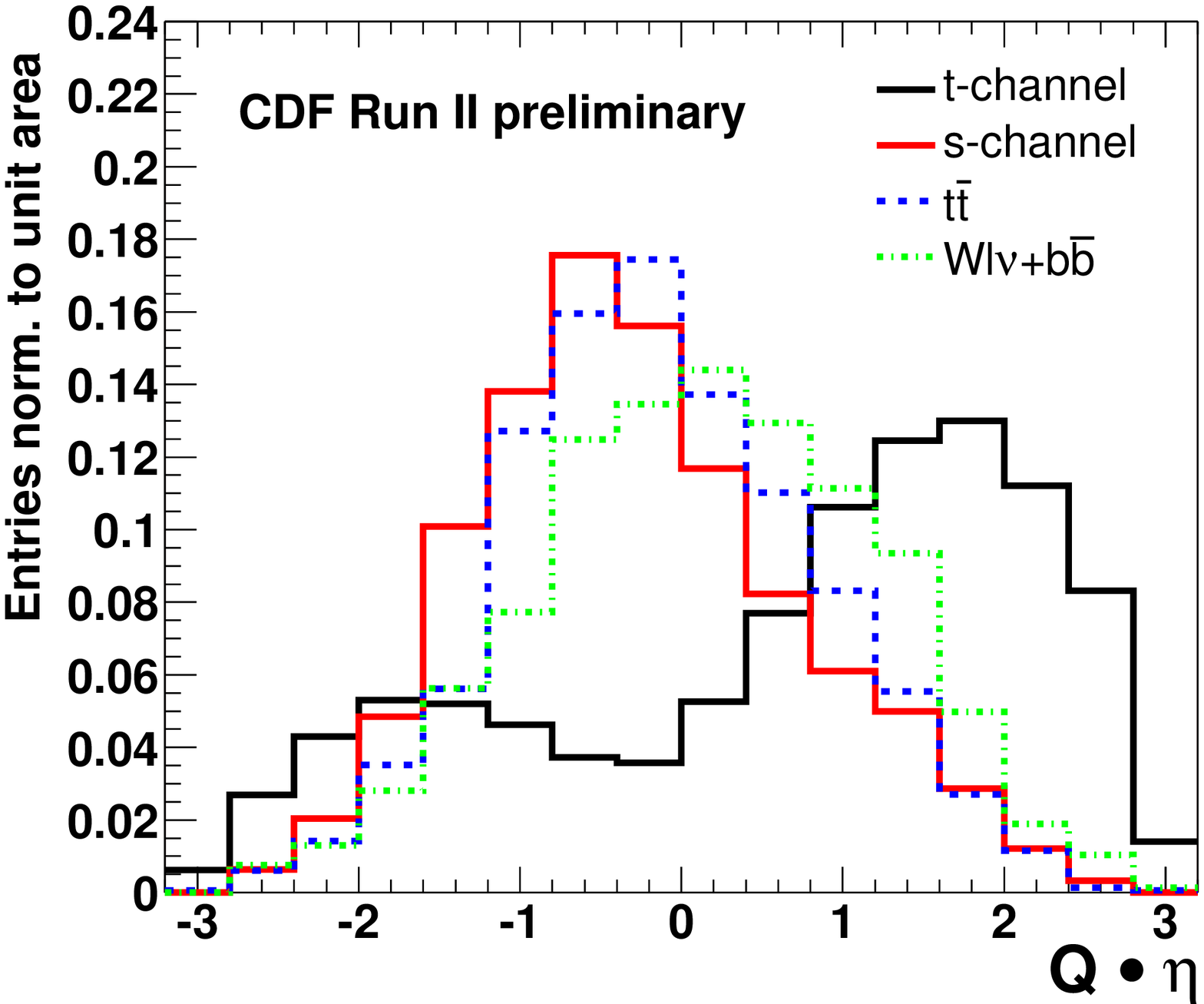}
    }
  }
 \caption{\it
    Expected distributions of $H_{T}$ (left) and $Q\cdot\eta$ (right) for single top
and associated backgrounds at CDF.
    \label{fig:ttemp} }
\end{figure}

CDF has searched for $s$- and $t$-channel production of single top at Run~II by
selecting events with an energetic lepton, large \met, and two jets with at least
one b-tag.  These events are ``sandwiched'' between \ttbar\ events at higher jet
multiplicities and W+jets events at lower jet multiplicities, making a counting
experiment extremely challenging.  Therefore the strategy adopted is to fit
the data to a distribution with some discriminating power between single top,
\ttbar, and other backgrounds.  The scalar $E_{T}$ sum of all objects in the
event ($H_{T}$) is one such distribution.  The $t$-channel provides an additional
handle for separating out single top production; the direction of the forward jet
in the event is highly correlated with whether a t or $\mathrm{\bar{t}}$ is produced,
and hence with the sign of the charged lepton in the event.  Therefore the product
of the pseudorapidity $\eta$ of the forward jet and the sign of the lepton charge
($Q\cdot\eta$)
tends to peak at large positive values for real $t$-channel single top production.
The expected distributions of these two discriminating variables are shown in
Figure~\ref{fig:ttemp}.

\begin{figure}[htbp]
  \centerline{\hbox{ 
    \includegraphics[width=0.45\textwidth]{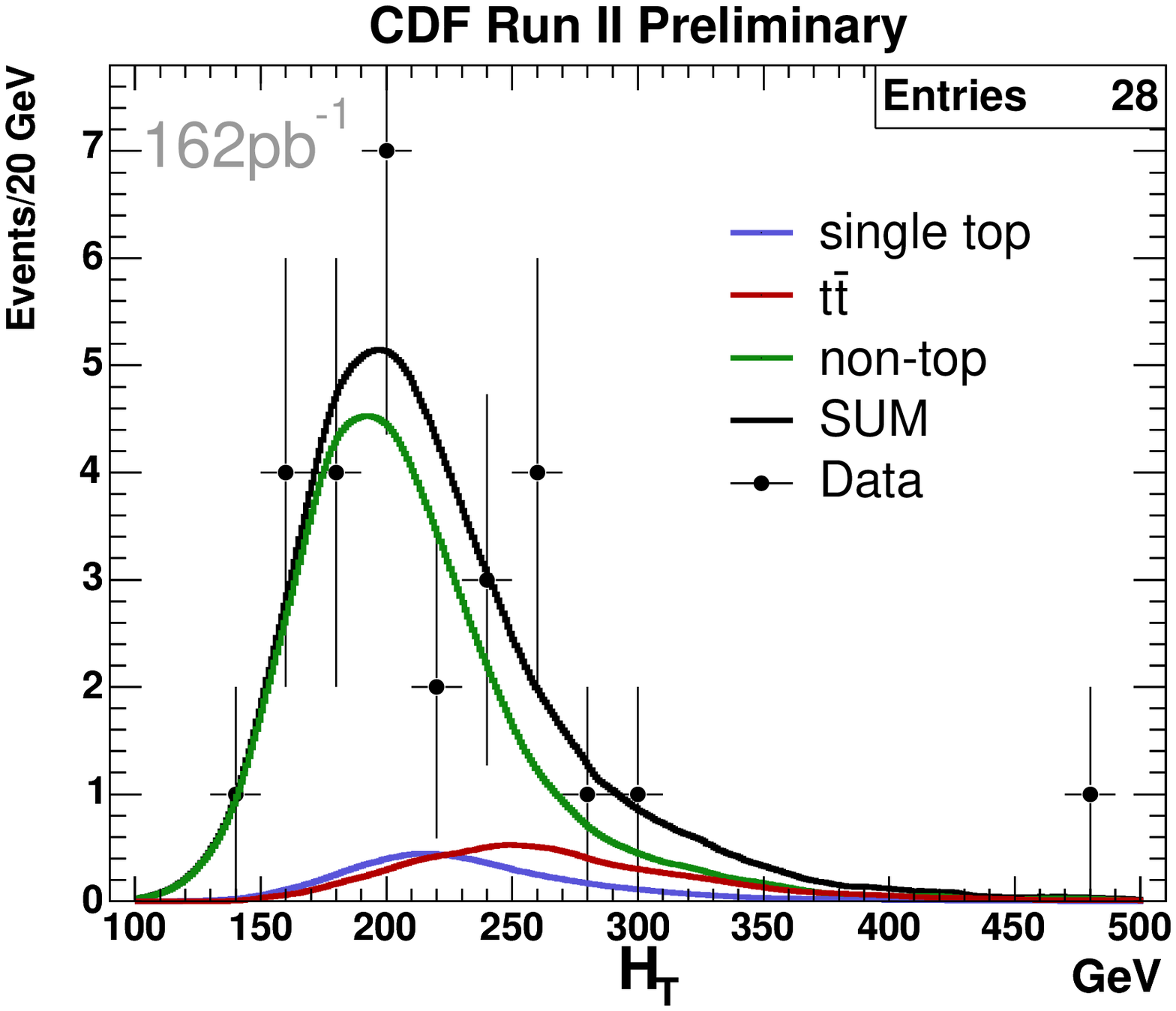}
    \hspace{0.3cm}
    \includegraphics[width=0.45\textwidth]{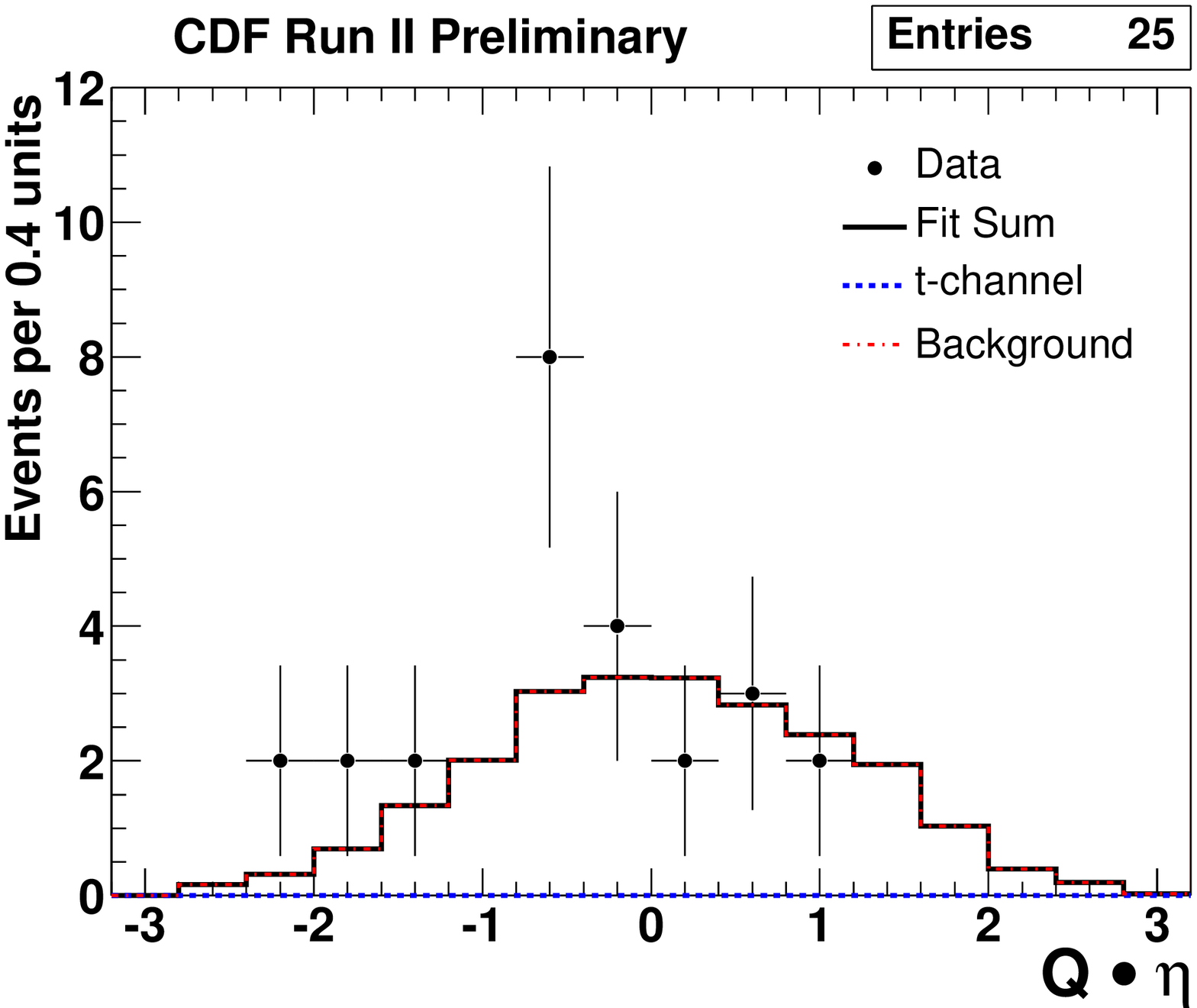}
    }
  }
 \caption{\it
    Distributions of $H_{T}$ (left) and $Q\cdot\eta$ (right) in the CDF Run~II data.
Also shown are the results of the fits to the templates shown in Figure~\ref{fig:ttemp}.
    \label{fig:tfit} }
\end{figure}

The CDF Run~II data are shown along with the fitted contributions from single top
and background in Figure~\ref{fig:tfit}.  No significant single top signal is
evident.  From the $H_{T}$ fit, a 95\% CL upper limit on the total $s$- and
$t$-channel cross section can be set at 13.7~pb.  Similarly, the $Q\cdot\eta$
fit yields an upper limit of 8.5~pb on the $t$-channel cross section alone.
Current projections estimate that 2~\invfb\ of data will be necessary to
report observation of single top, if there is no enhancement of the cross
section from new physics.

\section{Conclusions}
The Run~II top physics program is well underway at the Tevatron.  Many measurements
have been re-established and there is much activity directed toward improving
upon the techniques developed for Run~I.  Measurements of production cross sections,
mass, and decay properties do not yet reveal any deviations from Standard Model
expectations for the top quark; however, uncertainties on these measurements are
still large and leave plenty of room for new physics processes to be uncovered.
The anticipated large Run~II datasets will allow for far more stringent tests of
the top sector in the years to come.

\section{Acknowledgments}
The author would like to thank all the people in the CDF and D0 top
groups (too many to name) whose work went into the results presented,
and especially Arnulf Quadt for the assistance he provided in pulling
together the D0 results.


\begin{thebibliography}{99}

\bibitem{disco}
F.~Abe {\it et al}, Phys. Rev. {\bf D50}, 2966 (1994) \\
F.~Abe {\it et al}, Phys. Rev. Lett. {\bf 74}, 2626 (1995) \\
S.~Abachi {\it et al}, Phys. Rev. Lett. {\bf 74}, 2632 (1995)

\bibitem{tt_xsec}
R.~Bonciani {\it et al}, Nucl. Phys. {\bf B529}, 424 (1998),
updated in \linebreak arXiv:hep-ph/0303085 \\
N.~Kidonakis and R.~Vogt, Phys. Rev. {\bf D68}, 114014 (2003)

\bibitem{lepew}
The LEP EW Working Group
(\url{http://lepewwg.web.cern.ch/LEPEWWG/})

\bibitem{d0mass}
V.M.~Abazov {\it et al}, Nature {\bf 429}, 638 (2004)

\bibitem{bandh}
M.~Barnett and L.~Hall, Phys. Rev. Lett. {\bf 77}, 3506 (1996)

\bibitem{singlet_xsec}
B.W.~Harris {\it et al}, Phys. Rev. {\bf D66}, 054024 (2002)

\end{thebibliography}
\end{document}